# Does visual experience influence arm proprioception and its lateralization? Evidence from passive matching performance in congenitally-blind and sighted adults.

Najib M. ABI CHEBEL[1], Florence GAUNET[2], Pascale CHAVET[1],

Christine ASSAIANTE[3], Christophe BOURDIN[1] & Fabrice R. SARLEGNA[1]*

[1] Aix Marseille Univ, CNRS, ISM, Marseille, France

[2] Aix Marseille Univ, CNRS, LPC, Marseille, France

[3] Aix Marseille Univ, CNRS, LNC, Marseille, France

***Correspondence should be addressed to:** Fabrice R. Sarlegna
Address: Institute of Movement Sciences, 163 av. de Luminy, CP 910, 13009 Marseille, France.
E-mail: fabrice.sarlegna@univ-amu.fr

**Funding:** This work was supported by the Fondation des Aveugles de Guerre, Carnot Star Institute, Carnot Cognition Institute, NeuroMarseille Institute and the French government under the Programme « Investissements d'Avenir », Initiative d'Excellence d'Aix-Marseille Université via A*Midex funding (AMX-19-IET-004), and ANR (ANR-17-EURE-0029).

**Declarations of interest:** none.

**Number of Figures**: 3. **Number of Tables:** 1. **Number of Multimedia:** 0.
**Number of Words:** Abstract - 164 words; Introduction - 1242 words; Discussion – 1132 words.

**Abstract**

In humans, body segments' position and movement can be estimated from multiple senses such as vision and proprioception. It has been suggested that vision and proprioception can influence each other and that upper-limb proprioception is asymmetrical, with proprioception of the non-dominant arm being more accurate and/or precise than proprioception of the dominant arm. However, the mechanisms underlying the lateralization of proprioceptive perception are not yet understood. Here we tested the hypothesis that early visual experience influences the lateralization of arm proprioceptive perception by comparing 8 congenitally-blind and 8 matched, sighted right-handed adults. Their proprioceptive perception was assessed at the elbow and wrist joints of both arms using an ipsilateral passive matching task. Results support and extend the view that proprioceptive precision is better at the non-dominant arm for blindfolded sighted individuals. While this finding was rather systematic across sighted individuals, proprioceptive precision of congenitally-blind individuals was not lateralized as systematically, suggesting that lack of visual experience during ontogenesis influences the lateralization of arm proprioception.

Laterality – Elbow – Wrist – Upper limb.

**Introduction**

Proprioception describes the perception of body and limb position based on proprioceptors, specialized mechanosensory neurons that convey information about the stretch and tension experienced by muscles, tendons, skin and joints (Woo et al., 2015). The information encoded by proprioceptors contributes to action control as well as to conscious perception of body configuration. The importance of proprioception for action control has been highlighted by the motor impairments observed in individuals with impaired proprioception due to experimental vibration (Goodman & Tremblay, 2021; Verschueren et al., 1999) or due to a sensory neuropathy (for reviews, Cole & Paillard, 1995; Desmurget et al., 1998; Jayasinghe et al., 2021). Individuals with sensory neuropathy have massive proprioceptive impairments which are associated with massive motor impairments and in particular increased variability in performance (Sarlegna et al., 2010; Spencer et al., 2005). In such 'proprioceptively-deafferented' individuals, and in healthy individuals, arm motor performance depends on vision (Blouin et al., 1993; Elliott & Chua, 1996; Smeets et al., 2006; Spencer et al., 2005). It has also been suggested that arm proprioception, which needs to be fine-tuned throughout the lifespan, can be calibrated by vision (for reviews, Cressman & Henriques, 2011; Desmurget et al., 1998; Ostry & Gribble, 2016; see also Gaunet & Thinus-Blanc, 1996; Gori et al., 2010; Mirdamadi & Block, 2020).

In line with the idea that vision can calibrate proprioception, one would logically expect that a visual impairment results in impaired proprioception, consistent with the general-loss hypothesis (Cappagli et al., 2017). While there are few studies on the consequences of a visual impairment on proprioception, the general-loss hypothesis is supported by findings of visual deprivation resulting in impaired motor behavior in animals (Fine & Park, 2018; Hein & Held, 1967; Wiesel & Hubel, 1965). One way to explore the influence of visual deprivation on perception and action in humans is to study blind individuals. Previous work in blind revealed motor impairments but also perceptual impairments such as impaired auditory spatial localization (Cappagli et al., 2017; Gori et al., 2014; Zwiers et al., 2001) and impaired haptic orientation judgment (Gori et al., 2010). Proprioceptive reproduction was reported to be impaired in congenitally-blind individuals (Cappagli et al., 2017; but see also Nelson et al., 2018). However, the use of voluntary movements in the study of Cappagli et al. (2017) precluded a pure evaluation of how blindness influences the sense of proprioception considering the known role of efferent signals in joint position sense (Bhanpuri et al., 2013; Gandevia et al., 2006). In a study assessing proprioceptive perception of passive arm movements, Fiehler et al. (2009) reported that arm proprioception was impaired in congenitally-blind individuals except when they received early training in orientation and mobility capacities. Overall, the proprioceptive impairments found in blind individuals support the idea that vision can be helpful in calibrating proprioception.

Studies of individuals with a sensory impairment have provided ample evidence that the lack or loss of sensory information in one modality can result in compensatory plasticity. However, it is natural to question the cross-modal, compensatory hypothesis when considering the interactions between vision and proprioception given the central role that vision may play in calibrating proprioception. Support for the compensatory hypothesis in blind individuals can be found in previous work which reported supranormal memory (Amedi et al., 2003; Pasqualotto et al., 2013), supranormal auditory perception (Finocchietti et al., 2023; Lessard et al., 1998; Röder et al., 1999) and supranormal tactile perception (Van Boven et al., 2000; Wong et al., 2011). According to the compensatory hypothesis, a visual impairment could result in an improved proprioceptive perception. Consistent with this idea, Gaunet and Rossetti (2006) reported that arm pointing performance can be better in blind individuals than in blindfolded sighted participants (see also Jones 1972). In addition, Yoshimura et al. (2010) provided evidence that arm movement control relies more on proprioception in blind individuals than in blindfolded sighted individuals. Because pointing or matching an unseen voluntary movement can rely on proprioceptive as well as efferent signals, studying one's ability to match the unseen position of a passively-moved joint is a method of choice to specifically assess proprioception (Cressman & Henriques, 2011; Fuentes & Bastian, 2010; Goble & Brown, 2008b; Mirdamadi & Block, 2020; Oh et al., 2023; Ostry & Gribble, 2016; Velay et al., 1989). Using such a passive method, Ozdemir et al. (2013) reported that ankle proprioception was more accurate in blind individuals than in sighted participants. However, to the best of our knowledge, there is no report of a proprioceptive assessment for a specific upper-limb joint in blind individuals. It thus remains unclear how a visual impairment influences joint proprioception in the upper limb.

The goal of the present study was to determine, in humans, the influence of early visual experience on upper-limb joint proprioception. One issue is that proprioception is not uniform across arm joints, and across arms, as it depends on several neurophysiological and biomechanical factors. For example, it has repeatedly been shown that proprioceptive estimates are more accurate and/or precise (less variable) for the elbow joint than for the wrist joint (Abi Chebel et al., 2022; Sevrez & Bourdin, 2015; Sturnieks et al., 2007). There is also evidence that proprioceptive estimates of an upper-limb joint are more accurate and/or precise for the non-dominant arm than for the dominant arm (Abi Chebel et al., 2022; Goble & Brown, 2008a, 2008b). Indeed, asymmetry in performance between arms is a prominent feature of human behavior (Adamo & Martin, 2009; Elliott & Chua, 1996; Sainburg, 2016; Serrien et al., 2006). However, the mechanisms underlying the lateralization of arm proprioception are still not well understood.

The present study was designed to specifically test the hypothesis that lateralization of arm proprioception may be influenced by early visual experience. This idea stems from the suggestion that lateralization of manual aiming is linked to visual and proprioceptive feedback processing. Indeed, motor control of the dominant arm has been reported to rely more on visual feedback processing, while motor control of the non-dominant arm has been reported to rely more on proprioceptive feedback processing (for reviews, Elliott & Chua, 1996; Goble & Brown, 2008a; Sainburg, 2016). This may be due to asymmetries in development, during which the non-dominant arm could learn to rely more on proprioception and the dominant arm could learn to rely more on vision, which plays a predominant role in motor development (Assaiante et al., 2014; Assaiante & Amblard, 1995; Gaunet & Thinus-Blanc, 1996). If vision critically influences lateralization in humans, it is thus possible that lack of visual experience may influence the lateralization of motor and perceptual functions. In line with this idea, reduced lateralization of language (oral understanding) and emotion processing has been found in congenitally-blind individuals (Gamond et al., 2017; Lane et al., 2017; Röder et al., 2000). However, little is known about the link between visual experience and lateralization of arm proprioception.

In the present study, we tested the hypothesis that early visual experience influences proprioceptive lateralization by comparing passive proprioceptive perception of blindfolded sighted and congenitally-blind individuals. Considering that the putative influence of visual experience may depend on the age at onset of blindness, only congenitally-blind individuals were recruited. As proprioception is known to vary across body parts, we assessed proprioceptive perception at the elbow and wrist joints of both arms. With regard to the reduced lateralization hypothesis in blind individuals, we predicted that proprioceptive perception is more precise for the non-dominant arm compared to the dominant arm for sighted individuals but not for congenitally-blind individuals.

**Methods**

*Participants*

Congenitally-blind participants (see Table 1) were recruited over a period of 5 years from various associations for the blind located in Marseille (see Acknowledgments) and through snowball sampling. Sighted controls were recruited from Aix-Marseille University and Marseille city. Inclusion criteria for all participants included being right-handed, 18 years old or over, and free from diabetes or any cognitive or upper-limb sensorimotor deficit. Before the beginning of the experiment, all participants were provided with a consent form which was read to the blind participants and signed by every participant. This research protocol was approved by the national ethics committee CERSTAPS (IRB00012476-2020-03-06–60) and conducted in line with the Declaration of Helsinki.

To determine the minimum sample size required for this study, we performed a statistical power analysis using G*Power software (version 3.1.9.6; Kiel University, Kiel, Germany). We based our sample size calculation on the effect size found in Abi Chebel et al. (2022) on the interlimb differences in proprioception for sighted right-handed participants: for a F-test, 2x2x2 ANOVA (number of measurements per participant = 24) with a partial $\eta^2$ of 0.743, the minimum required sample size was estimated to be 4. Although we did our best to recruit more congenitally-blind participants, we had strict inclusion criteria and could test 'only' 8 individuals, a sample which corresponds to the sample size used in several similar studies (Finocchietti et al., 2023; Lessard et al., 1998; Röder et al., 1999). We recruited 8 sighted individuals who were matched for age and sex.

All 16 participants had a strong right-hand dominance, as determined with the 10-item version of the Edinburgh handedness inventory (Appendix II in Oldfield, 1971). The congenitally-blind group consisted of 3 females and 5 males [M (mean) ± standard deviation age = 43.5 ± 13.4 years (min - max: 21 - 61 years); M laterality quotient = 78.8 ± 15.5%; see Table 1]. Four participants were totally blind since birth (#1, 6, 7, and 8 in Table 1). The other four participants were able to perceive shadows and contrasts at birth and became totally blind between 16 and 25 years of age (#2, 3, 4 and 5 in Table 1). The sighted group consisted of 3 females and 5 males [M age = 42.9 ± 16.8 years (min - max: 21 - 69 years); M laterality quotient = 83.8 ± 19.4%] with no history of visual impairment. There were no significant differences in age and laterality quotient between the two groups, as revealed by two independent t-tests (t = 0.1, p = 1.0; t = -1.4, p = 0.2, respectively).

| Blind participant | Etiology | Independence in orientation and mobility* | Laterality quotient (%) | Age | Sex | Occupation |
|---|---|---|---|---|---|---|
| 1 | Congenital glaucoma | Low | 60 | 21 | Male | Student |
| 2 | Congenital eye malformation | Very low | 100 | 26 | Male | Job seeker |
| 3 | Congenital glaucoma | High | 60 | 44 | Male | Association volunteer |
| 4 | Congenital glaucoma | Moderate | 80 | 45 | Male | Technology instructor for blind |
| 5 | Leber congenital amaurosis | High | 100 | 49 | Female | Engineer |
| 6 | Congenital glaucoma | High | 70 | 50 | Female | Association volunteer |
| 7 | Congenital glaucoma | Moderate | 80 | 53 | Female | Association volunteer |
| 8 | Congenital glaucoma | High | 80 | 61 | Male | Piano tuner and repairer |

**Table 1: Description of the congenitally-blind participants.**
* Very low: assistance to go out; Low: familiar paths only; Moderate: familiar paths mainly; High: new and familiar paths.

*Experimental setup*

The setup for this study was similar to that used in Abi Chebel et al. (2022), and inspired by several studies (Fuentes & Bastian, 2010; Goble & Brown, 2008b; Sevrez & Bourdin, 2015; Velay et al., 1989). Seated participants placed their arm in an exoskeleton and grasped a handle with their hand. Their forearm was wrapped to a lever with fabric fasteners. For each participant, the exoskeleton was adjusted to align its mechanical rotation axes with the wrist and elbow rotation axes. This setup allowed near-frictionless movement at the wrist (hand movement) and elbow (forearm movement) in the horizontal plane at chest height.

Joint rotations were recorded with precision potentiometers (linear, 10 kΩ, Vishay) mounted at the pivot points of the apparatus. Each potentiometer was connected to an analog-to-digital converter connected to a computer. To record participants' verbal responses, a microphone (Scarlett CM25 MkIII, Focusrite, High Wycombe, UK) was positioned at the mouth level of each participant with a 'magic arm'. All signals were synchronized and sampled at 1 kHz using the LabView virtual instrument (National Instruments Corporation, Austin, TX, USA).

*Experimental procedures and conditions*

The experimenter presented the apparatus to all participants and helped congenitally-blind participants to explore it with their hands. Each participant was seated comfortably, given oral instructions, and blindfolded, except for one blind participant (#6 in Table 1) who had two eye prostheses and refused to be blindfolded. While one joint was being tested, the ipsilateral non-tested joint was immobilized by locking the corresponding part of the exoskeleton, and the contralateral arm rested on the participant's thigh. Each participant was tested on four experimental conditions, with each condition corresponding to one of the four tested joints (the right and left wrists and elbows).

Figure 1 illustrates the ipsilateral passive matching task. For each trial, the experimenter slowly moved the participant's body segment, according to the experimental condition (e.g., the left hand for the non-dominant wrist condition, or the right forearm for the dominant elbow condition). The body segment was moved to a random angle within a standardized start zone, located between 125 and 135° of flexion (0° corresponding to the elbow - shoulder axis) for the elbow and between 5 to 15° of flexion (0° corresponding to the wrist - elbow axis) for the wrist. From that angle, the experimenter (always NAC) moved the participant's body segment to the reference at a slow speed (<5°/s; as in Abi Chebel et al. (2022). As shown in Figure 1, the reference angle was set to 100° of elbow flexion and 30° of wrist extension, as in Goble et al. (2006) and Adamo and Martin (2009) respectively. Once the reference angle was reached, the experimenter stabilized the joint at that angle for 8 seconds to allow participants to focus on, and memorize, the current joint position. Once this memorization phase was completed, the experimenter slowly returned the participants' body segment to a random position in the start zone. Once the start zone was reached, the experimenter slowly extended the participant's body segment toward the memorized reference angle. Participants had to say 'Top' when they believed that their joint angle corresponded to the memorized reference angle, marking the end of the trial. Extreme ranges of motion were avoided and movement speed was controlled below 5°/second using visual feedback on a computer screen, as in Abi Chebel et al. (2022).

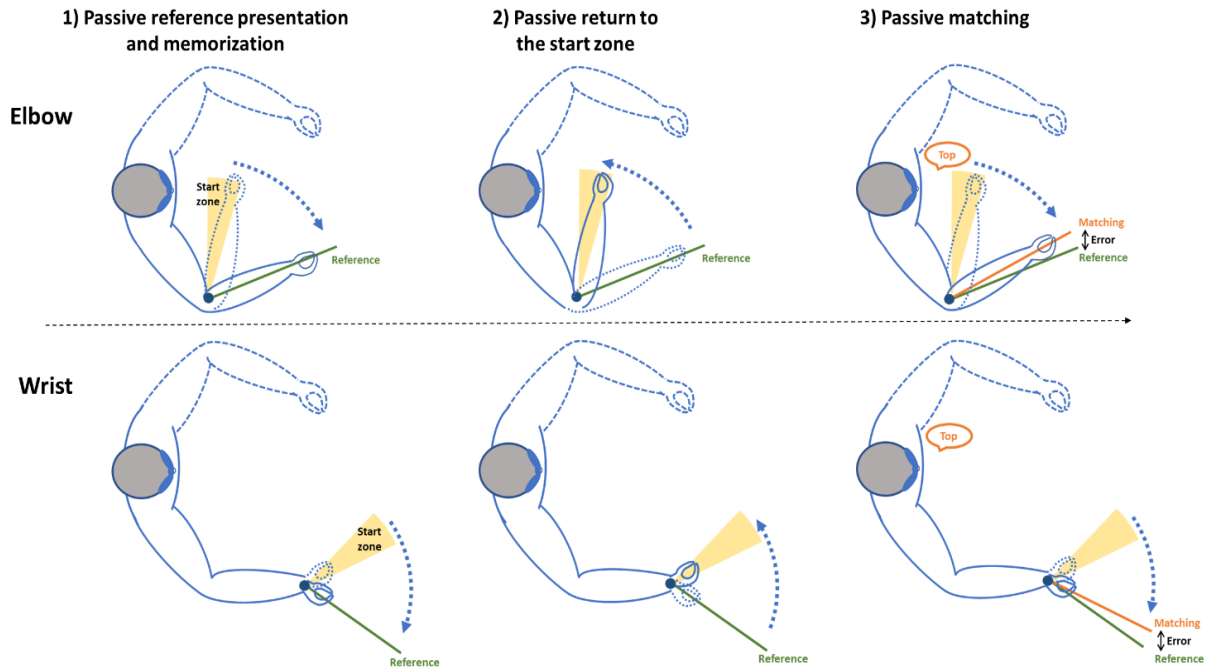

**Figure 1: Ipsilateral passive matching task.**

Top view of a participant in a right elbow condition (top panel) and right wrist condition (bottom panel).

First (left panels), the tested arm segment was slowly positioned by the experimenter in the start zone (orange, with arm segment drawn in broken lines) before being slowly moved to the reference (green line) and held there for memorization (8 seconds). Then (middle panels), the arm segment was slowly moved back from the reference (green line, with arm segment drawn in broken lines) to the start zone (orange). Then (right panels), the arm segment was slowly moved from the start zone (orange, with arm segment drawn in broken lines) toward the memorized reference. Participants had to say 'Top' when they perceived that the joint angle matched the reference angle.

The opposite arm is represented in broken lines to illustrate that the joints of both arms were tested, in distinct conditions, in a symmetrical way.

For each experimental condition, we conducted a session that consisted of 6 consecutive trials. A resting time of one to two minutes was given between each session. For participants' comfort, we tested arms in blocks, i.e. at first the elbow and wrist joints of one side (right or left) and then the other side. Since we counterbalanced the order of the joints and sides, one of the eight possible orders was randomly assigned to each participant. Participants were not given knowledge of results, as in Goble & Brown (2010).

*Data analysis*

Data were analyzed with Matlab (Mathworks R2020b) and Excel (Microsoft Office Professional Plus 2019) routines. To describe the participants' matching behavior across the six trials per experimental condition, five measures (in degrees) were computed:

- The mean absolute error was the mean of the 6 absolute differences between the reported joint angle and the reference angle (as in Goble & Brown, 2008b). The mean absolute error allowed focusing on the error amplitude, irrespective of its direction.
- The mean signed error was the mean of the 6 differences between the reported joint angle and the reference angle (as in Goble & Brown, 2008b). It was specifically useful to determine the directional bias of the matching performance. Positive mean signed errors were assigned to an overshoot of the reference angle. Negative mean signed errors were assigned to an undershoot of the reference angle.
- The variable signed error was the standard deviation around the mean signed error, reflecting the precision of proprioceptive perception which has been highlighted as an important aspect of performance (Abi Chebel et al., 2022; Fuentes & Bastian, 2010; Goble & Brown, 2008b; Oh et al., 2023).

- The variable absolute error was calculated as the standard deviation around the mean absolute error (as in (Abi Chebel et al., 2022; Sevrez & Bourdin, 2015).

Mixed-design analyses of variance (ANOVA) were conducted on performance variables to determine differences between groups (Visual experience: Congenitally blind, Sighted controls) and within participants (repeated measures) as a function of the factors Arm (Non-dominant, Dominant) and/or Joint (Wrist, Elbow) as well as their interactions. Statistical analyses were performed with STATISTICA (Version 7.1) and JASP (Version 0.16.3).

All raw data were normally distributed, as verified with the Kolmogorov-Smirnov method. Significance was set at $p < 0.05$. Post hoc comparisons were performed based on Newman–Keuls method and partial eta squared were reported as a measure of effect size where appropriate. Raw and processed data are available on the Open Science Framework public repository (https://osf.io/6angk/).

**Results**

*Mean errors in proprioceptive perception*

The mean absolute error of matching performance was analyzed to first focus on error amplitude. A 2x2x2 mixed-design ANOVA [Visual experience (Congenitally blind, Sighted control) x Arm (Non-dominant, Dominant) x Joint (Elbow, Wrist)] did not show any significant main effects of visual experience ($F(1,14) = 0.8$, $p = 0.4$, partial $\eta^2 = 0.05$), arm ($F(1,14) = 3.4$, $p = 0.1$, partial $\eta^2 = 0.2$), or joint ($F(1,14) = 0.8$, $p = 0.4$, partial $\eta^2 = 0.05$). The ANOVA also did not show any significant interactions between visual experience and arm ($F(1,14) = 2.5$, $p = 0.1$, partial $\eta^2 = 0.1$), visual experience and joint ($F(1,14) = 1.4$, $p = 0.3$, partial $\eta^2 = 0.1$), and arm and joint ($F(1,14) < 0.01$, $p = 0.1$, partial $\eta^2 < 0.01$), nor a significant double interaction ($F(1,14) = 0.3$, $p = 0.6$, partial $\eta^2 = 0.02$). Overall, mean absolute errors averaged approximately 4° (M congenitally blind = 4.2 ± 1.9°; M sighted control = 3.7 ± 1.5°).

Mean signed error was analyzed to take into account error direction. A 2x2x2 ANOVA did not show any significant main effects of visual experience ($F(1,14) = 0.6$, $p = 0.4$, partial $\eta^2 = 0.04$), arm ($F(1,14) = 2.1$, $p = 0.2$, partial $\eta^2 = 0.1$), or joint ($F(1,14) = 0.3$, $p = 0.6$, partial $\eta^2 = 0.02$). The ANOVA also did not show any significant interaction between visual experience and arm ($F(1,14) = 0.02$, $p = 0.9$, partial $\eta^2 < 0.01$), visual experience and joint ($F(1,14) = 0.1$, $p = 0.8$, partial $\eta^2 < 0.01$), and arm and joint ($F(1,14) = 0.2$, $p = 0.7$, partial $\eta^2 = 0.01$), nor a significant double interaction ($F(1,14) = 0.8$, $p = 0.4$, partial $\eta^2 = 0.05$). Overall, mean signed errors were relatively small for both groups of participants (M congenitally blind = 0.0 ± 3.7°; M sighted control = -1.1 ± 3.1°).

*Variable errors in proprioceptive perception*

It is well known that in addition to central tendency measures, dispersion measures are useful to understand properties of processes and provide information on sample heterogeneity. We first analyzed variable absolute error with a 2x2x2 ANOVA which only revealed a significant arm effect ($F(1,14) = 9.1$, $p < 0.01$, partial $\eta^2 = 0.4$). The variable absolute error at the non-dominant arm (M = 2.5 ± 1.2°) was significantly smaller than at the dominant arm (M = 3.3 ± 1.3°). There was no significant main effects of visual experience ($F(1,14) = 1.0$, $p = 0.3$, partial $\eta^2 = 0.06$) or joint ($F(1,14) = 0.7$, $p = 0.4$, partial $\eta^2 = 0.05$), nor any significant interactions between visual experience and arm ($F(1,14) = 1.8$, $p = 0.2$, partial $\eta^2 = 0.1$), visual experience and joint ($F(1,14) = 2.7$, $p = 0.1$, partial $\eta^2 = 0.2$), and arm and joint ($F(1,14) = 0.0$, $p = 1.0$, partial $\eta^2 < 0.01$), nor a significant double interaction ($F(1,14) = 0.0$, $p = 0.9$, partial $\eta^2 < 0.01$).

A 2x2x2 ANOVA on variable signed error revealed a significant arm effect ($F(1,14) = 8.1$, $p = 0.01$, partial $\eta^2 = 0.4$) and a significant interaction effect between arm and visual experience ($F(1,14) = 5.7$, $p = 0.03$, partial $\eta^2 = 0.3$). This interaction is illustrated in Figure 2. Newman-Keuls' post-hoc tests showed that for the sighted group, the variable signed error was smaller at the non-dominant arm compared to the dominant arm (Figure 2A; M non-dominant = 2.6 ± 1.0°; M dominant = 4.7 ± 1.6°, $p = 0.01$). In contrast, variable signed errors did not significantly differ between arms for the congenitally blind participants (Figure 2B-C; M non-dominant = 4.1 ± 2.0°; M dominant = 4.3 ± 1.9°, p = 0.7). Post-hoc analysis also revealed that the variable signed error was significantly smaller at the non-dominant arm of the sighted group compared to the dominant arm of the congenitally-blind group (p = 0.03). The variable signed error tended to be smaller at the non-dominant arm of the sighted group compared to the non-dominant arm of the congenitally-blind group (p = 0.054).

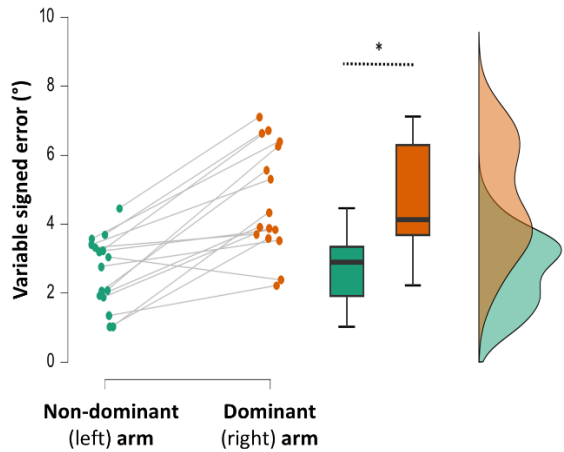
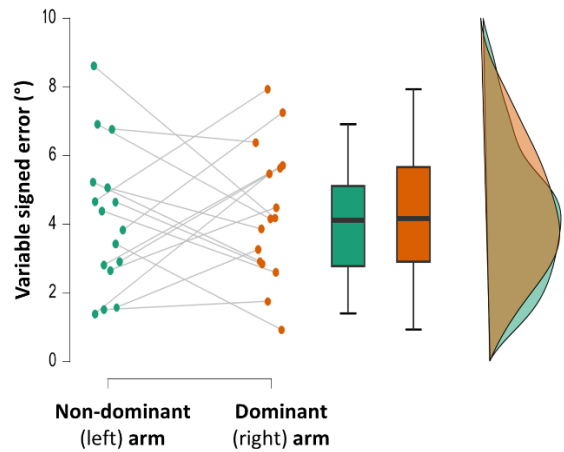
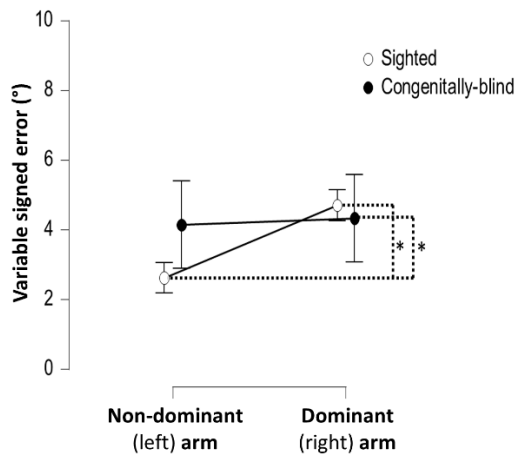
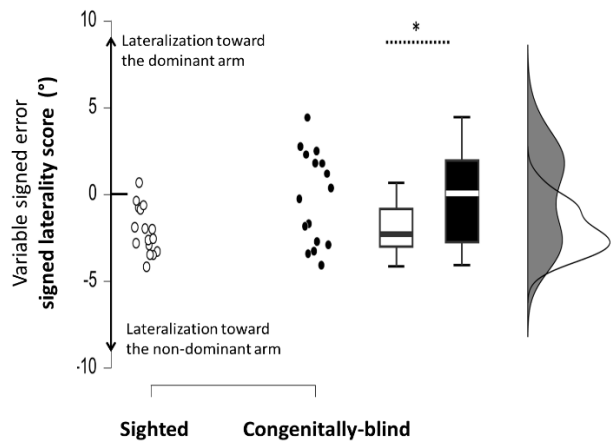

**Figure 2.** Mean variable signed error as a function of visual experience and arm.

A-B) Data for the sighted and congenitally-blind groups, respectively.

C-D) Data for the two experimental groups.

Panels A), B), and D) present dots for data of each participant, box and whisker plots (with minimum, maximum, median, first and third inter-quartile values) and data distribution.

Error bars in panel C) represent the 95 % confidence interval around the mean.

No significant main effect of visual experience ($F(1,14) = 1.2$, $p = 0.3$, partial $\eta^2 = 0.1$) or joint ($F(1,14) = 0.3$, $p = 0.6$, partial $\eta^2 = 0.02$) were found on the variable signed error, nor a significant interaction between arm and joint ($F(1,35) = 0.2$, $p = 0.62$, partial $\eta^2 < 0.01$), nor a significant double interaction ($F(1,14) = 0.1$, $p = 0.7$, partial $\eta^2 < 0.01$). A significant interaction between visual experience and joint was found on the variable signed error ($F(1,14) = 4.9$, $p = 0.04$, partial $\eta^2 < 0.01$) with no significant differences in post-hoc tests.

In summary, there was no significant difference between arms in the congenitally-blind group whereas the variable signed error was smaller at the non-dominant arm compared to the dominant arm in the sighted group. Figure 2A shows that such non-dominant arm advantage in the precision of proprioceptive perception was noticeable on most participants in the sighted group. To further assess lateralization differences between sighted and congenitally-blind individuals, we computed a laterality score by subtracting the variable signed error of the non-dominant arm to that of the dominant arm. A 2x2 mixed-design ANOVA [Visual experience (Congenitally blind, Sighted control) x Joint (Elbow, Wrist)] on such laterality score revealed a significant effect of visual experience ($F(1,14) = 5.7$, $p = 0.03$, partial $\eta^2 = 0.29$) but no significant main effect of joint ($F(1,14) = 0.1$, $p = 0.8$, partial $\eta^2 < 0.01$) and no significant interaction $F(1,14) = 0.1$, $p = 0.7$, partial $\eta^2 < 0.01$). Figure 2D shows that the laterality score of the sighted group ($M = -2.1 \pm 1.4°$) was smaller than that of the congenitally-blind group ($M = -0.2 \pm 2.7°$). Since a negative laterality score corresponds to a proprioceptive advantage for the non-dominant arm, these findings support the idea of a greater lateralization toward the non-dominant arm for the sighted group compared to the congenitally-blind group. A 2x2 mixed-design ANOVA [Visual experience (Congenitally blind, Sighted control) x Joint (Elbow, Wrist)] on the absolute value of the laterality score for the variable signed error did not reveal any significant effect of visual experience (F(1,14) = 0.2, p = 0.7, partial $\eta^2$ = 0.01), or joint (F(1,14) = 0.3, p = 0.6, partial $\eta^2$ = 0.02), and no significant interaction (F(1,14) < 0.01, p = 0.9, partial $\eta^2$ < 0.01). Overall, these analyses suggest that the direction of lateralization differed between groups but the amount of lateralization did not significantly differ between groups.

The non-dominant arm advantage in variable signed error was rather systematic across sighted individuals, in contrast to congenitally-blind individuals. To determine whether proprioceptive perception in congenitally-blind individuals was linked to other variables, we used linear correlation analyses and found a significant negative correlation between the variable signed error of the non-dominant arm and the laterality quotient in congenitally-blind individuals.

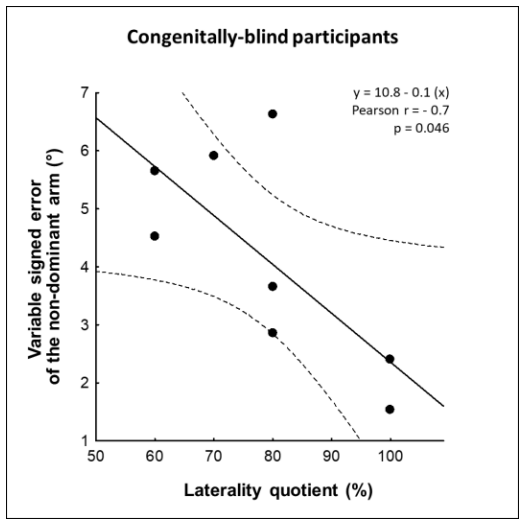

**Figure 3.** Correlation between the variable signed error of the non-dominant arm and the laterality quotient in congenitally-blind individuals.

**Discussion**

The current study aimed to determine whether early visual experience influences upper-limb proprioception and its lateralization by comparing blindfolded sighted and congenitally-blind individuals in a passive matching task. Consistent with our hypotheses, we found that proprioception was more precise for the non-dominant arm compared to the dominant arm in sighted individuals. This finding was rather systematic while in contrast, proprioception in congenitally-blind individuals did not significantly differ between arms. This suggests that lifelong lack of visual experience alters the typical asymmetry of arm proprioceptive precision typically observed in sighted individuals.

*Non-dominant arm advantage in arm proprioceptive precision for sighted individuals*

In the present study, we obtained evidence for the well-documented finding of a proprioceptive perception advantage for the non-dominant arm in sighted individuals (Abi Chebel et al., 2022; Goble & Brown, 2008b, 2010). Such asymmetry was found in both types of variable errors (signed and absolute). Both accuracy and precision values in the present study were consistent with those in the literature on the elbow and wrist joints (Adamo & Martin, 2009; Fuentes & Bastian, 2010; Goble & Brown, 2008b; Sevrez & Bourdin, 2015). In the present study, we did not find significant differences in proprioception between the elbow and wrist joints, and overall, mean errors in proprioceptive perception did not significantly differ across upper-limb joints, as previously observed (Tripp et al., 2006). The most sensitive measure was the variability of errors in proprioceptive perception, as variable errors differed between (dominant and non-dominant) arms and (sighted and congenitally-blind) groups. In the last few decades, the emergence of the Bayesian framework of sensory integration has sparked interest in the precision (variability) of arm position sense (Smeets et al., 2006; Van Beers et al., 2002). Studying the accuracy as well as the precision has proven to be useful in better characterizing the proprioceptive sense across multiple joints and populations for instance (Fiehler et al., 2009; Fuentes & Bastian, 2010; Oh et al., 2023). When proprioceptive perception was assessed in a previous study on healthy adults, analysis of variable errors was critical as the main finding was that proprioception of the non-dominant arm was more precise compared to the dominant arm (Abi Chebel et al., 2022).

It has been suggested that proprioceptive asymmetries may be related to asymmetries in manual preferences, performances, and/or control processes (for reviews, Elliott & Chua, 1996; Goble & Brown, 2008a). For instance, Bagesteiro and Sainburg (2003) found more effective proprioceptively-mediated responses to unexpected load perturbations for the non-dominant arm, supporting the idea of a specialization of each arm for distinct proprioceptive and visual control processes (for reviews, Jayasinghe et al., 2021; Sainburg, 2016). In this framework, non-dominant arm advantages may be associated with more precise proprioceptive estimates of limb position and movement, which would be consistent with the better proprioceptive precision of sighted participants observed in the present study as well as in previous work.

*Early visual experience influences proprioceptive lateralization*

It has been suggested that lateralization of functions may be influenced by visual experience. In the present study, the better precision of proprioceptive perception for the non-dominant arm was rather systematic across (blindfolded) sighted participants. The same pattern was found only for few congenitally-blind participants, suggesting that the lack of early visual experience during ontogenesis prevents the improvement of proprioceptive precision for a specific arm across the population. These findings are consistent with the view that visual experience influence the lateralization of neural networks, as supported by differences between congenitally-blind and sighted individuals for some functions such as sentence understanding (Lane et al., 2017; Röder et al., 2000) and emotional processing (Gamond et al., 2017). Overall, our findings and other findings support the view that early visual experience, or the lack thereof, leads to changes in brain structures and functions.

Previous studies have suggested a right-hemisphere dominance in proprioceptive perception of sighted individuals (Ben-Shabat et al., 2015; Chilvers et al., 2021; Goble et al., 2012; Naito et al., 2005; Strong et al., 2023). Here, we speculate that blindness may be associated with changes in proprioceptive lateralization due to differences in lateralization of proprioceptive networks in sensorimotor and cerebellar areas. To test this hypothesis, future research could investigate the neural bases of proprioceptive perception and their lateralization in sighted and congenitally-blind individuals. One may see that brain organization is more variable in the blind compared to the sighted, as suggested by recent work on brain connectivity (Sen et al., 2023).

In congenitally-blind individuals, the precision of proprioceptive perception for the non-dominant arm was linked to the laterality quotient, i.e., hand preferences in daily activities. This suggests that in the absence of visual experience, lateralization of arm proprioception is influenced by the lateralization of arm use, and possibly by the daily activities and the type of sensory (auditory, tactile…) feedback used by blind individuals. This is consistent with a study of Fiehler et al. (2009) which reported that early training in orientation and mobility for congenitally-blind individuals can benefit arm proprioception so that it is as good as for sighted individuals, suggesting that arm proprioception can be fine-tuned in different ways.

Overall, further work is necessary to determine how proprioception is lateralized in congenitally-blind participants. In the present study, participants were well matched in terms of age, sex and manual laterality. One possibility is that for a given task, the difference in proprioceptive lateralization between sighted and congenitally-blind participants results from a complex interaction between visual experience, lateralization, task specificity and individual characteristics (Goble & Brown, 2008a; Lane et al., 2017; Sainburg, 2016; Serrien et al., 2006). Indeed, individual characteristics have been shown to influence proprioception: for instance, Fiehler et al. (2009) reported proprioceptive differences between congenitally-blind participants with or without early orientation and mobility training. In a developmental framework, it would be interesting to assess sensory and motor skills with children and adults to determine the interactions between sensory experience, motor experience and proprioceptive perception.

An obvious limitation of the present study is the relatively small sample size. Over a five-year period (including the coronavirus pandemic), we could not test more right-handed congenitally-blind volunteers who had no associated pathology. This prevented us from assessing possible influences of factors such as etiology, orientation and mobility training or habits, specific skills or activity. Despite this limitation, the current study has implications for our understanding of the relationship between vision and proprioception. The finding that early visual experience may play a crucial role in the lateralization of proprioceptive precision during development supports the idea that vision contributes to the calibration of proprioception. This is consistent with the idea of using visual feedback for the development of technological aids or rehabilitation protocols for individuals with proprioceptive impairments, and consistent with previous work which has highlighted the dependency on vision for proprioceptively-impaired individuals (Blouin et al., 1993; Cole & Paillard, 1995; Spencer et al., 2005).

**Acknowledgements**

We would like to extend our deep appreciation to all the participants of our study for their invaluable contributions. We would also like to thank Patrick Sainton, Frank Buloup, and Thelma Coyle for their technical assistance. We are also grateful to the associations for the visually-impaired for their support and collaboration throughout the study (Association de Réadaptation et de Réinsertion pour l'Autonomie des Déficients Visuels; Union Nationale des Aveugles et Déficients Visuels; Association Valentin Haüy; IRSAM; Fédération des Aveugles et Handicapés Visuels de France – Union Provençale des Aveugles et Amblyopes des Cannes Blanches; Association Sports et Loisirs des Aveugles et Amblyopes).

# References

Abi Chebel, N. M., Roussillon, N. A., Bourdin, C., Chavet, P., & Sarlegna, F. R. (2022). Joint Specificity and Lateralization of Upper Limb Proprioceptive Perception. *Perceptual and Motor Skills*, *129*(3), 431–453. https://doi.org/10.1177/00315125221089069

Adamo, D. E., & Martin, B. J. (2009). Position sense asymmetry. *Experimental Brain Research*, *192*(1), 87–95. https://doi.org/10.1007/s00221-008-1560-0

Amedi, A., Raz, N., Pianka, P., Malach, R., & Zohary, E. (2003). Early "visual" cortex activation correlates with superior verbal memory performance in the blind. *Nature Neuroscience*, *6*(7), 758–766. https://doi.org/10.1038/nn1072

Assaiante, C., & Amblard, B. (1995). An ontogenetic model for the sensorimotor organization of balance control in humans. *Human Movement Science*, *14*(1), 13–43. https://doi.org/10.1016/0167-9457(94)00048-J

Assaiante, C., Barlaam, F., Cignetti, F., & Vaugoyeau, M. (2014). Body schema building during childhood and adolescence: A neurosensory approach. *Neurophysiologie Clinique*, *44*(1), 3–12. https://doi.org/10.1016/j.neucli.2013.10.125

Bagesteiro, L. B., & Sainburg, R. L. (2003). Nondominant arm advantages in load compensation during rapid elbow joint movements. *Journal of Neurophysiology*, *90*(3), 1503–1513. https://doi.org/10.1152/jn.00189.2003

Ben-Shabat, E., Matyas, T. A., Pell, G. S., Brodtmann, A., & Carey, L. M. (2015). The Right Supramarginal Gyrus Is Important for Proprioception in Healthy and Stroke-Affected Participants: A Functional MRI Study. Frontiers in Neurology, 6, 248. https://doi.org/10.3389/FNEUR.2015.00248

Bhanpuri, N. H., Okamura, A. M., & Bastian, A. J. (2013). Predictive modeling by the cerebellum improves proprioception. *Journal of Neuroscience*, *33*(36), 14301–14306. https://doi.org/10.1523/JNEUROSCI.0784-13.2013

Blouin, J., Bard, C., Teasdale, N., Paillard, J., Fleury, M., Forget, R., & Lamarre, Y. (1993). Reference systems for coding spatial information in normal subjects and a deafferented patient. *Experimental Brain Research*, *93*(2), 324–331. https://doi.org/10.1007/BF00228401/METRICS

Cappagli, G., Cocchi, E., & Gori, M. (2017). Auditory and proprioceptive spatial impairments in blind children and adults. *Developmental Science*, *20*(3), 1–12. https://doi.org/10.1111/desc.12374
27

Chilvers, M. J., Hawe, R. L., Scott, S. H., & Dukelow, S. P. (2021). Investigating the neuroanatomy underlying proprioception using a stroke model. Journal of the Neurological Sciences, 430. https://doi.org/10.1016/J.JNS.2021.120029

Cole, J., & Paillard, J. (1995). Living without Touch and Peripheral Information about Body Position and Movement: Studies with Deafferented Subjects. *The MIT Press*, 245–266.

Cressman, E. K., & Henriques, D. Y. P. (2011). Motor adaptation and proprioceptive recalibration. *Progress in Brain Research*, *191*, 91–99. https://doi.org/10.1016/B978-0-444-53752-2.00011-4

Desmurget, M., Pélisson, D., Rossetti, Y., & Prablanc, C. (1998). From eye to hand: Planning goal-directed movements. *Neuroscience and Biobehavioral Reviews*, *22*(6), 761–788. https://doi.org/10.1016/S0149-7634(98)00004-9

Elliott, D., & Chua, R. (1996). Manual asymmetries in goal-directed movement. In D. Elliott & E. A. Roy (Eds.), *Manual asymmetries in motor performance* (pp. 143–158).

Fiehler, K., Reuschel, J., & Rösler, F. (2009). Early non-visual experience influences proprioceptive-spatial discrimination acuity in adulthood. *Neuropsychologia*, *47*(3), 897–906. https://doi.org/10.1016/j.neuropsychologia.2008.12.023

Fine, I., & Park, J. M. (2018). Blindness and human brain plasticity. *Annual Review of Vision Science*, *4*, 337–356. https://doi.org/10.1146/annurev-vision-102016-061241

Finocchietti, S., Esposito, D., & Gori, M. (2023). Monaural auditory spatial abilities in early blind individuals. *I-Perception*, *14*(1), 1–9. https://doi.org/10.1177/20416695221149638

Fuentes, C. T., & Bastian, A. J. (2010). Where is your arm? Variations in proprioception across space and tasks. *Journal of Neurophysiology*, *103*(1), 164–171. https://doi.org/10.1152/jn.00494.2009

Gamond, L., Vecchi, T., Ferrari, C., Merabet, L. B., & Cattaneo, Z. (2017). Emotion processing in early blind and sighted individuals. *Neuropsychology*, *31*(5), 516–524. https://doi.org/10.1037/neu0000360

Gandevia, S. C., Smith, J. L., Crawford, M., Proske, U., & Taylor, J. L. (2006). Motor commands contribute to human position sense. *Journal of Physiology*, *571*(3), 703–710. https://doi.org/10.1113/jphysiol.2005.103093

Gaunet, F., & Rossetti, Y. (2006). Effects of visual deprivation on space representation: Immediate and delayed pointing toward memorised proprioceptive targets. *Perception*, *35*(1), 107–124. https://doi.org/10.1068/p5333
28